\documentclass[footinbib,aps,twocolumn,superscriptaddress,floatfix,prb]{revtex4-2}

\usepackage[utf8]{inputenc}
\usepackage{amssymb,amsmath,amsfonts,accents,amsthm}
\usepackage{mathtools}
\usepackage{braket}
\usepackage[colorlinks=true,citecolor=blue,linkcolor=blue,urlcolor=blue]{hyperref}
\usepackage{xcolor}

\usepackage[german,english]{babel}

\makeatletter
\def\bbl@set@language#1{%
	\edef\languagename{%
		\ifnum\escapechar=\expandafter`\string#1\@empty
		\else\string#1\@empty\fi}%
	\@ifundefined{babel@language@alias@\languagename}{}{%
		\edef\languagename{\@nameuse{babel@language@alias@\languagename}}%
	}%
	\select@language{\languagename}%
	\expandafter\ifx\csname date\languagename\endcsname\relax\else
	\if@filesw
	\protected@write\@auxout{}{\string\select@language{\languagename}}%
	\bbl@for\bbl@tempa\BabelContentsFiles{%
		\addtocontents{\bbl@tempa}{\xstring\select@language{\languagename}}}%
	\bbl@usehooks{write}{}%
	\fi
	\fi}
\newcommand{\DeclareLanguageAlias}[2]{%
	\global\@namedef{babel@language@alias@#1}{#2}%
}
\makeatother

\DeclareLanguageAlias{en}{english}
\DeclareLanguageAlias{English}{english}
\DeclareLanguageAlias{Englisch}{english}
\DeclareLanguageAlias{EN}{english}
\DeclareLanguageAlias{en-US}{english}
\DeclareLanguageAlias{de}{german}

\usepackage{graphicx}
\usepackage[export]{adjustbox}

\usepackage{xcolor}

\usepackage[most]{tcolorbox}
\tcbuselibrary{theorems}
\usepackage[capitalise]{cleveref}
\newtcbtheorem[	crefname={theorem}{Theorems},Crefname={theorem}{Theorem}]{Theorem}{Theorem}{
	colback=green!5,
	colframe=green!35!black,
	fonttitle=\bfseries,
	breakable,
}{thm}
\newtcbtheorem[	crefname={lemma}{Lemma},Crefname={lemma}{Lemma}]{Lemma}{Lemma}{
	colback=orange!5,
	colframe=orange!35!black,
	fonttitle=\bfseries,
	breakable,
}{lem}

\newtcbtheorem[	crefname={corollary}{Corollary},Crefname={corollary}{Corollary}]{Corollary}{Corollary}{
	colback=orange!5,
	colframe=orange!35!black,
	fonttitle=\bfseries,
	breakable,
}{cor}

\newtcbtheorem[
crefname={definition}{Definition},
Crefname={definition}{Definition}]
{Definition}{Definition}{
	colback=blue!5,
	colframe=blue!35!black,
	fonttitle=\bfseries,
}{def}

\newcommand{\ham}{H}

\newcommand{\optimizedFreq}{\overline{f}}

\newcommand{\component}[3]{\left(#1\right)_{#2,#3}}
\newcommand{\sMat}{S}

\newcommand{\matPowerIndex}{m}
\newcommand{\su}{a}
\newcommand{\sv}{b}

\usepackage{bbold}

\begin{document}
	\title{Equireflectionality and customized unbalanced coherent perfect absorption \\in asymmetric waveguide networks
 }

	\author{Malte Röntgen}
	\affiliation{%
		Zentrum für optische Quantentechnologien, Universität Hamburg, Luruper Chaussee 149, 22761 Hamburg, Germany
	}%
 	\affiliation{%
		Laboratoire d’Acoustique de l’Université du Mans, Unite Mixte de Recherche 6613, Centre National de la Recherche Scientifique, Avenue O. Messiaen, F-72085 Le Mans Cedex 9, France
	}%
 
	\author{Olivier Richoux}%
	\affiliation{%
		Laboratoire d’Acoustique de l’Université du Mans, Unite Mixte de Recherche 6613, Centre National de la Recherche Scientifique, Avenue O. Messiaen, F-72085 Le Mans Cedex 9, France
	}%
	
	\author{Georgios Theocharis}%
	\affiliation{%
		Laboratoire d’Acoustique de l’Université du Mans, Unite Mixte de Recherche 6613, Centre National de la Recherche Scientifique, Avenue O. Messiaen, F-72085 Le Mans Cedex 9, France
	}%

 	\author{Christian V. Morfonios}%
	\affiliation{%
		Zentrum für optische Quantentechnologien, Universität Hamburg, Luruper Chaussee 149, 22761 Hamburg, Germany
	}%

	\author{Peter Schmelcher}
	\affiliation{%
		Zentrum für optische Quantentechnologien, Universität Hamburg, Luruper Chaussee 149, 22761 Hamburg, Germany
	}%
	\affiliation{%
		The Hamburg Centre for Ultrafast Imaging, Universität Hamburg, Luruper Chaussee 149, 22761 Hamburg, Germany
	}%

        \author{Philipp del Hougne}
	\affiliation{%
		Univ Rennes, CNRS, IETR-UMR 6164, F-35000 Rennes, France
	}%

 	\author{Vassos Achilleos}%
	\affiliation{%
		Laboratoire d’Acoustique de l’Université du Mans, Unite Mixte de Recherche 6613, Centre National de la Recherche Scientifique, Avenue O. Messiaen, F-72085 Le Mans Cedex 9, France
	}%

	\begin{abstract}
    We explore the scattering of waves in designed asymmetric one-dimensional waveguide networks. 
    We show that the reflection between two ports of an asymmetric network can be identical over a broad frequency range, as if the network was mirror-symmetric, under the condition of so-called \emph{latent} symmetry between the ports.
    This broadband equireflectionality is validated numerically for acoustic waveguides and experimentally through measurements on microwave transmission-line networks. In addition, introducing a generalization of latent symmetry, we study the properties of an $N$-port scattering matrix $S$.
    When the powers of $S$ fulfill certain relations, which we coin \emph{scaled cospectrality}, the setup is guaranteed to possess at least one zero eigenvalue of $S$, so that the setup features coherent perfect absorption.
    More importantly, scaled cospectrality introduces a scaling factor which controls the asymmetry of the incoming wave to be absorbed. 
    Our findings introduce a novel approach for designing tunable wave manipulation devices in asymmetric setups. As evidenced by our acoustic simulations and microwave experiments, the generality of our approach extends its potential applications to a wide range of physical systems.
	\end{abstract}
	
	\maketitle
	
\section{Introduction}
Scattering of waves is unambiguously of fundamental importance in physics, finding applications in fields as diverse as high-energy physics \cite{Peskin2018IntroductionQuantumFieldTheory}, X-ray diffraction \cite{B.D.Cullity2001ElementsXRayDiffraction}, wave localization \cite{PSheng} or wave filtering \cite{Macleod2010ThinfilmOpticalFilters,Dokumaci2021DuctAcousticsFundamentalsApplications,SpringerHandbookAcoustics}.
In any scattering problem, the system's symmetries have a strong influence.
For a reflection symmetry---the simplest geometric symmetry---, for instance, waves sent into the system from the two opposite sides of the symmetry axis (or plane) act identically, thus the two reflection coefficents are strictly equal.

Here, on the other hand, we are interested in exploring the scattering properties of geometrically \emph{asymmetric} systems.
In particular, we focus on systems featuring the recently introduced  \emph{latent symmetry} \cite{Smith2019PA514855HiddenSymmetriesRealTheoretical,Kempton2020LAIA594226CharacterizingCospectralVerticesIsospectral,Rontgen2022LatentSymmetriesIntroduction}.
Such a symmetry is usually not visible in the original setup, but it becomes apparent after a suitable dimensional reduction, the so-called isospectral reduction \cite{Bunimovich2014IsospectralTransformationsNewApproach}.
What is interesting about a latent symmetry is its strong impact on the eigenmodes of the underlying system \cite{Rontgen2021PRL126180601LatentSymmetryInducedDegeneracies}.
A latent reflection symmetry, for instance, induces local parity on the eigenvectors of the underlying matrix $M$ describing the system (for instance, the Hamiltonian or the scattering matrix) \cite{Kempton2020LAIA594226CharacterizingCospectralVerticesIsospectral,Morfonios2021LAaiA62453CospectralityPreservingGraphModifications,Rontgen2023PRL130077201HiddenSymmetriesAcousticWave}.

Interestingly, the impact of a latent symmetry goes beyond the system's eigenvectors and manifests itself also in certain relations of the powers of the matrix $M$ \cite{Rontgen2021AQSymmetriesMatrixItsIsospectral}. For a latent reflection symmetry with respect to two sites $u,v$, the corresponding diagonal elements $\left(M^k \right)_{u,u}$ and $\left(M^k \right)_{v,v}$ of the matrix powers of $M$ are the same for all positive integers $k$ \cite{Kempton2020LAIA594226CharacterizingCospectralVerticesIsospectral}. 
On a fundamental level, this result is interesting, as it shows a deep connection between the powers of $M$ and its eigenvectors \cite{Morfonios2021LAaiA62453CospectralityPreservingGraphModifications}.

In the first part of this work, we carry the concepts of latent symmetry and of matrix powers to the realm of wave scattering.
We start by applying the concept of latent symmetry to design geometrically \emph{asymmetric} systems whose scattering properties carry the same traits as a reflection symmetric system, that is, broadband equireflectionality. We validate this intriguing property numerically in acoustic waveguide networks and experimentally in microwave transmission-line networks.

Then, in the second part of this work,  we explore the impact of relations in the matrix powers of a general scattering matrix. Specifically, we show that an $N$-port system can be designed to feature coherent perfect absorption (CPA) using \emph{scaled} matrix power relations. CPA implies the complete and irreversible transduction of the incident wave energy into other degrees of freedom such as heat, which is possible whenever the scattering matrix has a zero eigenvalue and the corresponding eigenvector is used as incident wavefront~\cite{ChongCPA}. This condition is valid irrespective of the complexity of the wave system, applying to simple~\cite{landy2008perfect} or disordered~\cite{f2020perfect} systems excited by a single channel as well as simple~\cite{wan2011time} or disordered~\cite{pichler2019random} systems excited by multiple channels. In asymmetric disordered systems, the necessary wavefront is generally very complex unless the system is optimized to impose CPA with a specific wavefront~\cite{delPRL2021}. Here, we show that a $3$-port system designed to feature scaled cospectrality will enable CPA with a prescribed imbalance of the corresponding wavefront that is controlled by tuning the associated scaling factor.

This paper is organized is follows. \Cref{sec:EquireflectionalScattering} is dedicated to the concept of equireflectional scattering, including theoretical and experimental verifications of the theory.
\Cref{sec:powersOfTheSMatrix} focuses on the powers of the scattering matrix of a generic $N$-port setup. In particular, we introduce scaled matrix power relations which can be directly related to CPA. We then focus on a simple three-port system, which we optimize to feature scaled cospectrality and hence CPA with custom imbalance of the CPA wavefront.
Finally, we conclude our work in \cref{sec:Conclusions}.

	\section{Scattering off latently symmetric networks} \label{sec:EquireflectionalScattering}

In the following, we will discuss how a latent reflection symmetry leads to equireflectional scattering.
After introducing this phenomenon by means of a simple example in \cref{sec:example}, we show in \cref{sec:UnderstandingEquireflectionality} how this phenomenon can be explained and designed; in  \cref{acoustic_simulations} and \cref{sec:microwave_experiments} we numerically and experimentally validate our findings with acoustic waveguide and microwave transmission-line networks, respectively.

\subsection{Setup and a first example} \label{sec:example}

	\begin{figure}[htb] 
		\centering
		\includegraphics[max width=\linewidth]{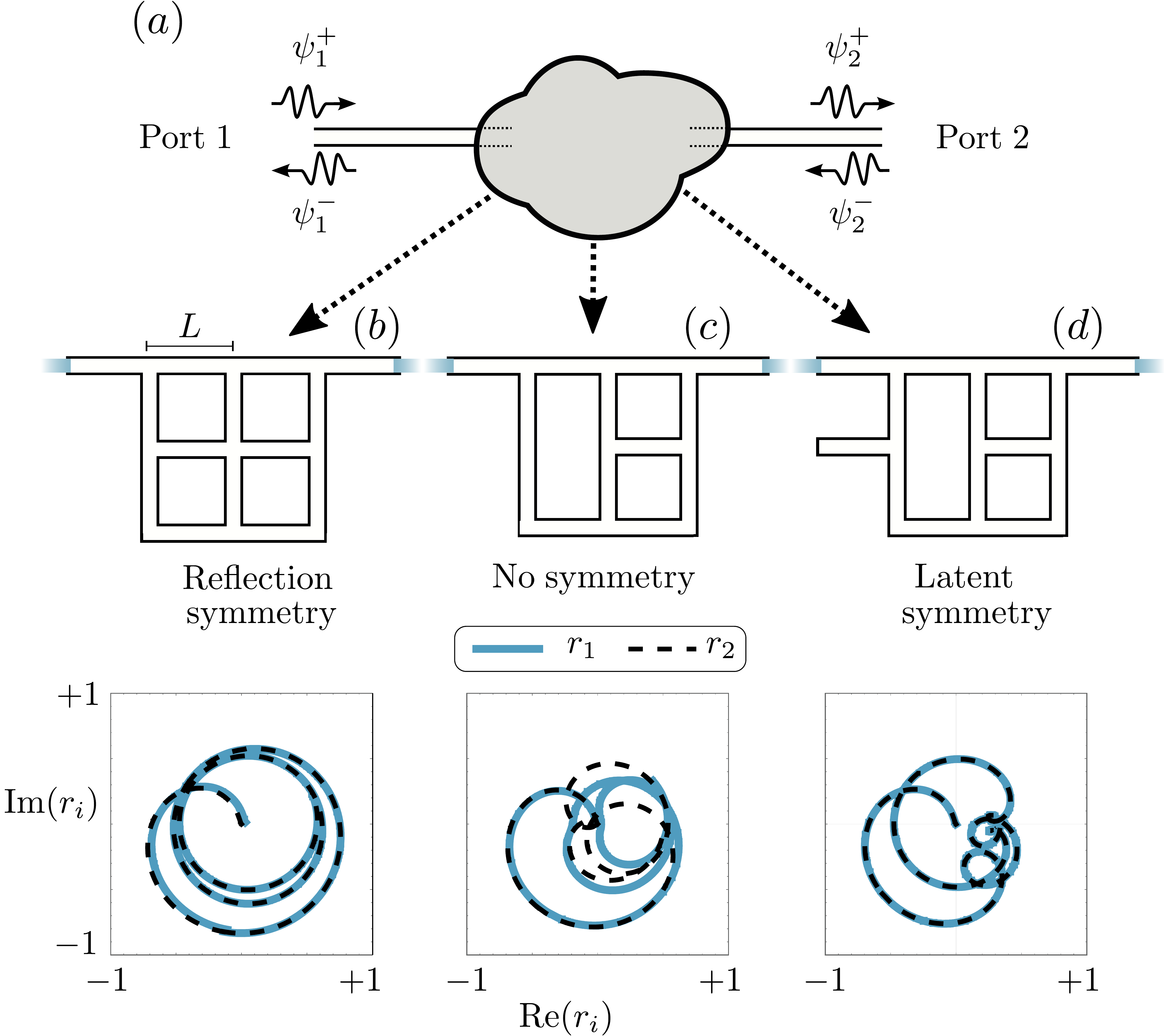}
		\caption{(a) A generic two-port scattering setup. (b-d): Different waveguide networks (upper panel) with their respective frequency-dependent reflection coefficients $r_{1}(f)$, $r_{2}(f)$ in the complex plane shown in the lower subpanel. These reflection coefficients have been computed using the method from \cite{Kottos2003JPAMG363501QuantumGraphsSimpleModel} for acoustic waves with visco-thermal losses (\cref{eq:transmissionInEachWaveguide,eq:lossesEquation} with loss-coefficient $\alpha = 3 \cdot 10^{-5}$), with a frequency $f$ between $0$ and $1000$ Hz. Each waveguide has a length $L=0.1m$ and a diameter $R=2mm$.
		}
		\label{fig:firstNetwork}
	\end{figure}

Consider a general reciprocal two port system, with only 
a single mode traveling within each lead/port  
as depicted in \cref{fig:firstNetwork}(a).
This problem is characterized by the following scattering matrix equation 
\begin{equation} \label{eq:prototypeSMatrix}
\begin{pmatrix} \psi_1^- \\ \psi_2^- \end{pmatrix} 
= \begin{pmatrix}
r_{1} & t \\
t & r_{2}
\end{pmatrix} \,
\begin{pmatrix} \psi_1^+ \\ \psi_2^+ \end{pmatrix} 
= S \begin{pmatrix} \psi_1^+ \\ \psi_2^+ \end{pmatrix},
\end{equation}
where $(\psi_1^-,\psi_2^-)^T$ and $(\psi_1^+,\psi_2^+)^T$ describe, respectively, the output and input waves. The off-diagonal elements of $S$ correspond to the transmission coefficients from left and right which, here, are equal due to reciprocity ($S = S^{T}$ is symmetric with respect to the diagonal).
The reflection coefficients from ports $1$ and $2$ are noted respectively $r_1$ and $r_2$. 

In this section, we consider networks of identical one-dimensional waveguide segments with the same length $L$, such as the ones shown in \cref{fig:firstNetwork}(b-d).
We note that networks of one-dimensional waveguides are known as ``quantum graphs'' and have been studied extensively in the last decades; an excellent introduction to the field is given in \cite{Berkolaiko2013186IntroductionQuantumGraphs}. Experimentally, they could be realized, for instance, in the form of microwave networks \cite{Hul2004PRE69056205ExperimentalSimulationQuantumGraphs,Hofmann2021PRE104045211SpectralDualityGraphsMicrowave} or networks of thin acoustic waveguides \cite{Coutant2021JoAP129125108TopologicalTwodimensionalSuSchriefferHeegerAnalog,Coutant2021PRB103224309AcousticSuSchriefferHeegerLatticeDirect,Rontgen2023PRL130077201HiddenSymmetriesAcousticWave}; in this paper, we will focus on the former for our experimental validation, while we focus on the latter for our numerical results.

 Let us now investigate the scattering off such a waveguide network, and focus on the  behavior of the reflection coefficients $r_1$ (from left) and $r_2$ (from right). To do so, here, we consider 1D acoustic waves of frequency $\omega=c_0 k$ where $c_0$ is the sound speed and $k$ the wavenumber. The reflection of the two-port system is given by the scattering matrix calculated using the 1D Helmholtz equation in each waveguide and the conservation of acoustic flux at the connections \cite{Richoux2020JPDAP53235101MultifunctionalResonantAcousticWave}.
 Note that this approximation is valid for acoustic propagation through the waveguides, assuming that $L \gg w$, where $L$ ($w$) is the length (width) of each waveguide.

 For a mirror-symmetric network, we expect \emph{equireflectionality}, that is, $r_{1} = r_2$, simply due to the total symmetry of the scattering problem. For illustration purposes, we show the two reflection coefficients of the network of \cref{fig:firstNetwork}(b) in the complex plane in the bottom panel, verifying equireflectionality. Notice that here we have considered the effect of homogeneously distributed losses since the reflection coefficients are inside the unit circle (see below for details). Evidently, such losses maintain the mirror symmetry.

 Breaking this mirror symmetry, as is the case in  \cref{fig:firstNetwork}(c), is expected to destroy equireflectionality. That is, $r_{1} \ne r_{2}$ except for some special frequencies where the two reflection coefficients coincide. This is clearly seen  in the bottom panel in \cref{fig:firstNetwork}(c). 
 On the other hand, for a special \emph{asymmetric} setup as the one in \cref{fig:firstNetwork}(d) we find that the system \emph{is}  equireflectional, \emph{i.e.}, $r_{1} = r_{2}$, as shown in the bottom panel, even in the presence of losses. This surprising result is not obtained by chance or coincidence. 
In fact, below, we show how such asymmetric equireflectional networks can be designed, and how their equireflectionality can be explained through the recently introduced concept of latent symmetry. 
	
		\begin{figure}[htb] 
		\centering
		\includegraphics[max width=\linewidth]{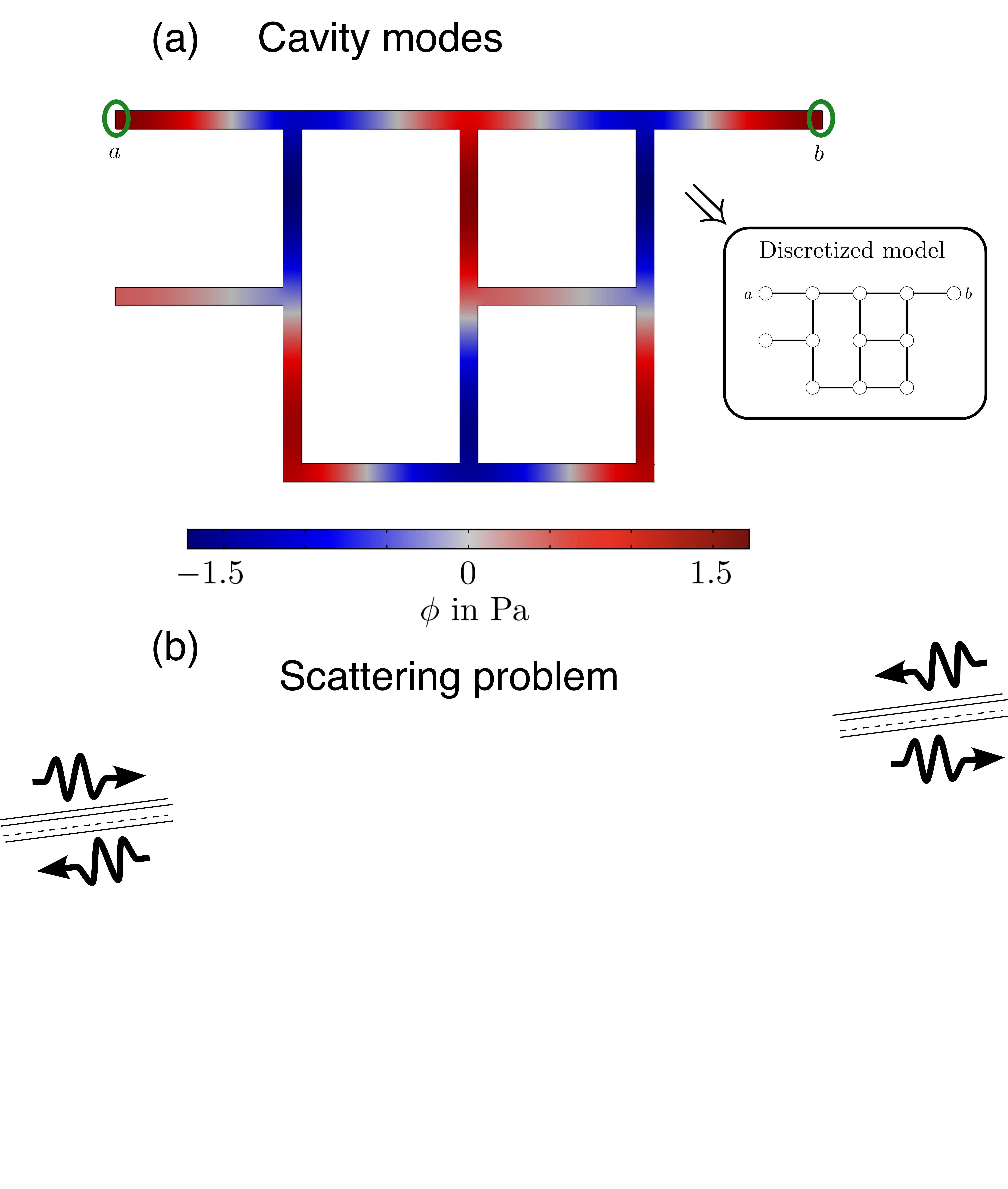}
		\caption{The network of \cref{fig:firstNetwork} (d), realized with thin, square-shaped acoustic waveguides of length $L=0.1m$ and side length $w=10mm$ (see \cref{acoustic_simulations} for details).
   (a) The 10th mode of the acoustic network with closed ends. The inset shows the corresponding discrete model (graph). (b) Schematic of the scattering process by the 2-port network. The acoustic pressure field here corresponds to a symmetric input (same amplitude and phase) from both ports at $f=625$Hz.}
		\label{fig:Mapping_To_A_Graph}
	\end{figure}
	
	\subsection{Review of latent symmetry in waveguide networks} \label{sec:latSymmMath}

\label{sec:UnderstandingEquireflectionality} To understand the highly symmetric scattering properties of the setup depicted in \cref{fig:firstNetwork}(d), we first need to study the eigenmodes of the corresponding cavity that is obtained by closing the connections to the leads, as shown in \cref{fig:Mapping_To_A_Graph}.
As we will show below, all these eigenmodes have the particular property of possessing \emph{point-wise} parity at the boundary points $a,b$ similarly to the case of a geometrically mirror symmetric setup.
That is, all eigenmodes $\phi$ feature $\phi(a) = \pm \phi(b)$; an example eigenmode demonstrating this property is shown in \cref{fig:Mapping_To_A_Graph} (a). 
We note that this property for acoustic waveguides has been studied in detail in Ref.~\cite{Rontgen2023PRL130077201HiddenSymmetriesAcousticWave} and here we review the basic results.

It can be  shown that the problem of finding the eigenmodes $\phi$ of the network is equivalent to the following generalized eigenvalue problem
	\begin{equation} \label{eq:gEVP}
		A\, \boldsymbol{\phi} = \cos(k L) B\, \boldsymbol{\phi}
	\end{equation}
	with $\cos(k L)$ being the eigenvalue, the $N$-dimensional eigenvector $\boldsymbol{\phi}$ denoting the values of the eigenmode $\phi$ at the $N$ endpoints of waveguides (see inset of \cref{fig:Mapping_To_A_Graph}(a) for details).
	The matrix $A$ describes the topology of the setup, with $A_{i,j} = 1$ if the endpoints $i,j$ are connected by a waveguide, and $A_{i,j} = 0$ otherwise.
	The matrix $B$ is diagonal, with $B_{i,i} = \sum_{j} A_{i,j}$.
    We then introduce the ``Hamiltonian'' $\ham = \sqrt{B}^{-1} A \sqrt{B}^{-1}$, which arises naturally from \cref{eq:gEVP} through the transformation $\mathbf{y} = \sqrt{B} \boldsymbol{\phi}$. 
    
We are then left with the eigenvalue problem
		\begin{equation} \label{eq:normalEVP}
		H \mathbf{y} = \cos(k L) \mathbf{y} \, .
	\end{equation}

Solving \cref{eq:normalEVP} gives us the natural frequencies of the eigenmodes for the network cavity, as well as the values of the field at the nodes. 

The claim is that, there are two different ways \cite{Rontgen2023PRL130077201HiddenSymmetriesAcousticWave} to obtain point-wise parity at some points $a,b$ for all eigenmodes. The first is through geometrical symmetry, that is, when the system is invariant under a symmetry operation---usually a reflection, as is the case for \cref{fig:firstNetwork} (a)---which maps the junctions $a$ and $b$ onto each other.
    The second and non-obvious case occurs when the setup is \emph{not} invariant under a geometrical operation that maps $a$ and $b$ onto each other. This is the case for the network of
\cref{fig:Mapping_To_A_Graph} (a) and \cref{fig:firstNetwork} (d).
In fact, the corresponding discrete network, as shown in the inset of \cref{fig:Mapping_To_A_Graph} (a), has been designed such that 
the  following relation of the matrix  powers of $\ham$ is fulfilled
	\begin{equation} \label{eq:matrixPowerRelations}
		\left(\ham^{\matPowerIndex} \right)_{\su,\su} = \left(\ham^{\matPowerIndex} \right)_{\sv,\sv}
	\end{equation}
	for $\matPowerIndex = 1,\ldots{},N$\,.
This has been found to ensure the pointwise parity of $a$ and $b$ for all the modes \cite{Rontgen2023PRL130077201HiddenSymmetriesAcousticWave}. 
 Note that \cref{eq:matrixPowerRelations} is automatically satisfied for a mirror symmetric setup. As we demonstrate in \cref{app:matrixPowerRelationsNeighbours}, the relations \cref{eq:matrixPowerRelations} can also be translated into a set of geometric rules---among others, the number of neighbors of $a,b$ have to be the same---that can be visually checked.
 Interestingly, the relations \cref{eq:matrixPowerRelations} have also a consequence on eigenvalues: If a matrix $\ham$ satifies these, then the two matrices $\ham\setminus a$ and $\ham \setminus \sv$---obtained from $\ham$ by removing the $\su$th (or $\sv$th) row and column---have the same eigenvalue spectra.
We thus say that $\su$ and $\sv$ are \emph{cospectral}.
 
 Before continuing, let us mention that the relations \cref{eq:matrixPowerRelations} lead to another very interesting consequence. By performing a certain dimensional reduction onto the two sites $\{a,b\}$---the so-called isospectral reduction \cite{Bunimovich2014IsospectralTransformationsNewApproach}, the resulting reduced system can be described by an effective Hamiltonian with a reflection symmetry. This explains  the point-wise parity of eigenvectors on $a,b$. Since this symmetry becomes in general only apparent after the dimensional reduction, we call the Hamiltonian $\ham$ \emph{latently reflection symmetric} \cite{Smith2019PA514855HiddenSymmetriesRealTheoretical,Rontgen2023PRL130077201HiddenSymmetriesAcousticWave,Rontgen2021PRL126180601LatentSymmetryInducedDegeneracies}.
    
\subsection{Equireflectionality in Acoustic Waveguide Networks}\label{acoustic_simulations}

Let us now investigate the scattering  properties of the system by connecting the points $a$ and $b$ to two waveguides with the same cross-section, as shown in \cref{fig:Mapping_To_A_Graph}(b). For this setup, the two incoming plane waves can be related to the outgoing waves through \cref{eq:prototypeSMatrix}. To find the corresponding scattering coefficients we first relate the acoustic pressure and its derivative between points $a$ and $b$ through a matrix involving the eigenmodes of the closed cavity. 
To do so, we use the integral formalism with an expression of the Green function of the closed cavity. Then, as explained  in detail in section IV of the supplemental material of \cite{Rontgen2023PRL130077201HiddenSymmetriesAcousticWave}, it can be shown that the former matrix is mirror symmetric due to the point-wise parity of the eigenmodes.
This leads to equireflerectionality $r_1=r_2$ of the corresponding scattering problem and thus the surprising  result of \cref{fig:firstNetwork} (d).

\begin{figure}[htb] 
		\centering
		\includegraphics[max width=\linewidth]{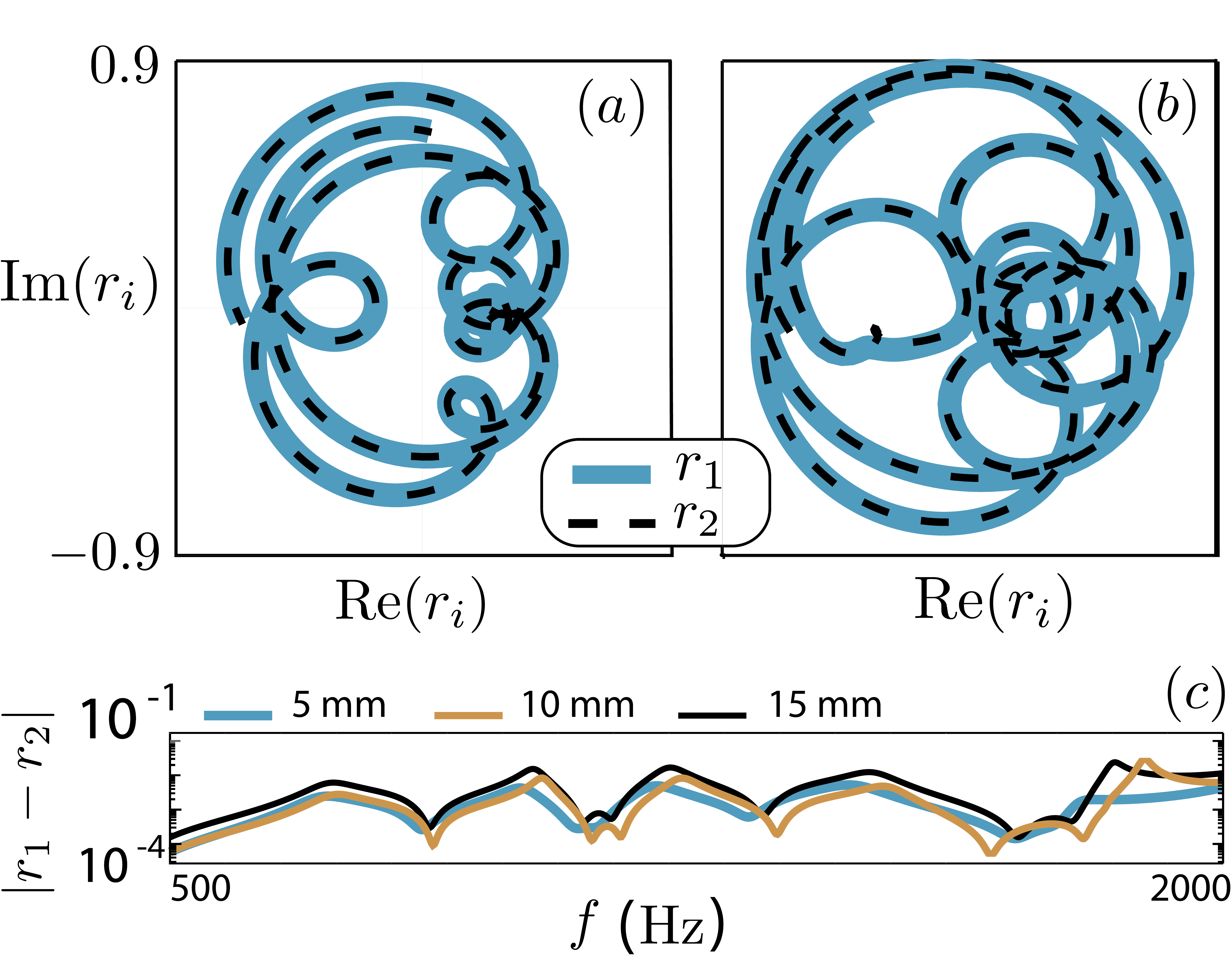}
		\caption{Results from $3D$ finite-element simulations of scattering off the structure of \cref{fig:firstNetwork}(d) using COMSOL, with visco-thermal losses modelled through the ``narrow region acoustics'' function. The waveguides are identical with a length of $L=0.1m$, and with a quadratic cross-section with side length $w$.
  (a) and (b) show a comparison of $r_{1}$ and $r_{2}$ for waveguide side lengths $w=5mm$ and $w=15mm$, respectively. The frequency range is between $0$ and $2000Hz$, which comprises a little more than the first $12$ eigenmodes of the underlying cavity. (c) Comparison of $|r_{1} - r_{2}|$ for three values of $w$.
		}
		\label{fig:frequencyBehavior}
	\end{figure}

So far, we have focussed on the case of thin waveguides.
When deviating from this limiting case of $w\ll L$, $3$D-effects at the junctions of the waveguides will play a role due to evanescent waves; their severity is, clearly, system-dependent (number of junctions, geometry of the connection etc).
We calculated the reflection coefficients for various values of  $w/L$ and show some characteristic results in \cref{fig:frequencyBehavior}. In particular we show the reflections $r_{1}(f),r_{2}(f)$ for the latently-symmetric scattering network of \cref{fig:firstNetwork} (d).
As can be seen, for this setup, $r_{1} = r_{2}$ holds quite well even when $w/L$ is far from the limiting value zero, although some deviations are seen in  \cref{fig:frequencyBehavior} (b). In \cref{fig:frequencyBehavior}(c) we show a more quantitative comparison of the behavior of these two reflections by showing the absolute value of the frequency-dependent difference $r_{1}(f) - r_{2}(f)$ for different sidelengths $w$.
In all of the calculations for \cref{fig:frequencyBehavior}, we have used viscothermal losses, which, as can be seen, do not alter equireflectionality.
Indeed, it turns out that this kind of losses does not at all impact the latent symmetry of the cavity or the equireflectionality of the (open) system (see \cref{app:lossySystems} for details).

\begin{figure}[htb] 
	\centering
	\includegraphics[max width=\linewidth]{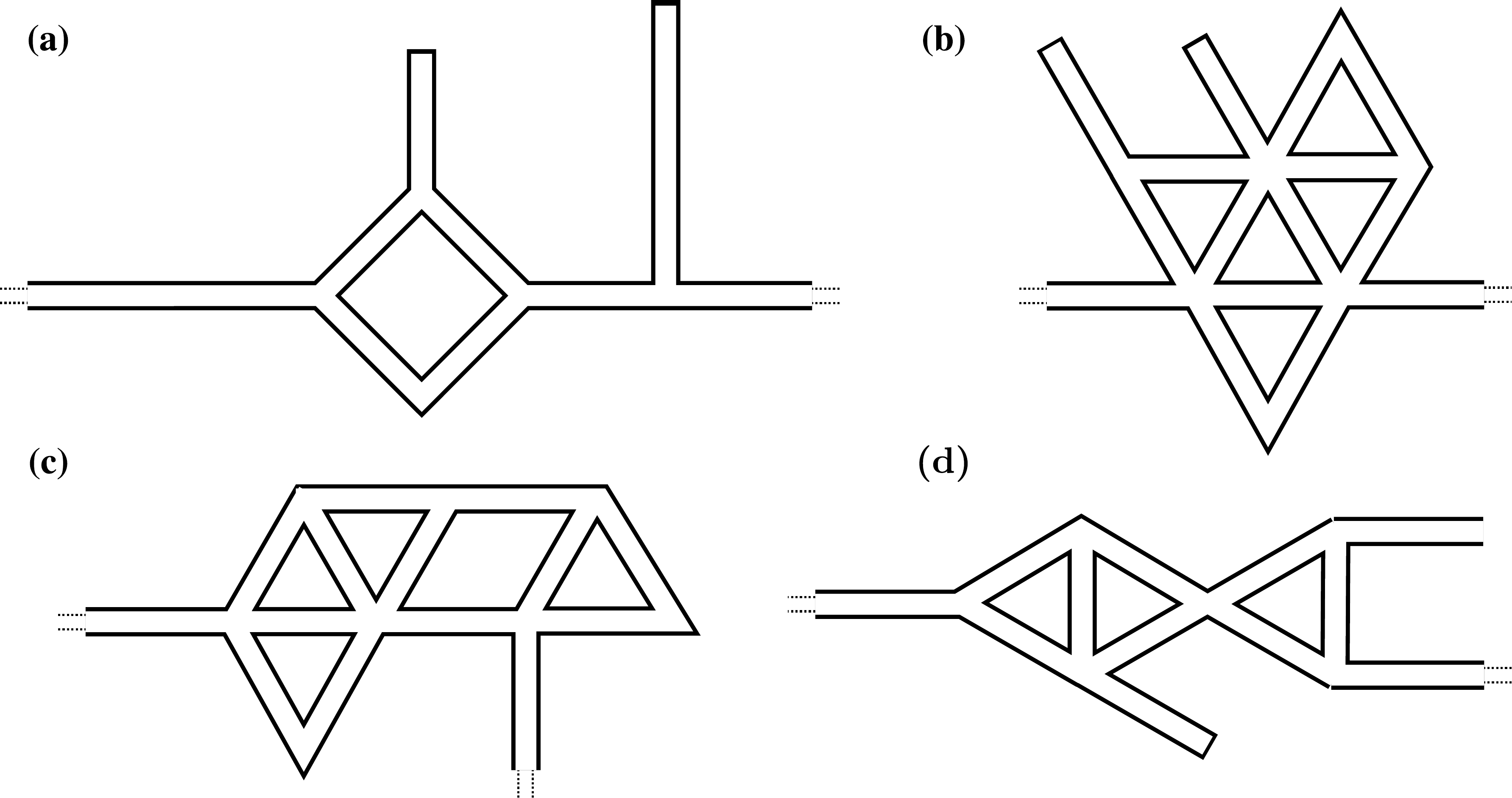}
	\caption{Different waveguide networks whose scattering matrix is equireflectional.
	}
	\label{fig:moreNetworks}
	\end{figure}

We stress that the equireflectional network used just above is not unique.
In fact, as long as networks are constructed under the assumption of one-dimensional (single mode) waveguides 
only the topology of the network, that is, which waveguide is connected to which, becomes important and not the actual geometry. Thus, one can use well-established tools from graph theory---such as the nauty-suite \cite{McKay2014JoSC6094PracticalGraphIsomorphismII}---to efficiently generate a large number of asymmetric graphs.
Each graph can be linked to a topology matrix $A$, from which we can construct the matrices $B$ and $\ham$.
Then we can test whether the corresponding waveguide network has a latent symmetry, that is, whether \cref{eq:matrixPowerRelations} is fulfilled. In this manner, a plethora of possible networks can be found (which are equireflectional). 
Some examples are shown in \cref{fig:moreNetworks}.
 
\subsection{Experimental Observation in Microwave Transmission-Line Networks}\label{sec:microwave_experiments}

As mentioned earlier, besides acoustic waveguide networks, microwave transmission-line networks lend themselves well to experimentally observe the discussed scattering signature of latent symmetry, namely equireflection.
Indeed, as can be easily shown \cite{Hul2004PRE69056205ExperimentalSimulationQuantumGraphs,Hofmann2021PRE104045211SpectralDualityGraphsMicrowave,AhmedMubarack2021TransmissionLinesQuantumGraphs}, the mathematical considerations in \cref{sec:latSymmMath} can be directly applied to transmission-line networks, if only we replace the pressure field $p$ with the voltage $U$. 
The only difference between the two platforms is that, in acoustics, a waveguide with a closed end (acoustic hard wall) features Neumann boundary conditions on that end; to achieve the same in transmission line networks, we need to make the corresponding cable \emph{open-ended}.
In summary, and taking into account this correspondence, the structures with latent symmetry from Fig.~\ref{fig:firstNetwork}(d) and Fig.~\ref{fig:moreNetworks} are directly applicable to a microwave realization.

We have hence built the networks from Fig.~\ref{fig:firstNetwork}(d) and Fig.~\ref{fig:moreNetworks}(a) using 50~cm-long coaxial cables and measured their scattering parameters with a vector network analyzer (VNA, Rhode \& Schwarz ZVA 67, 10 MHz – 67 GHz). The finite propagation delay in the junctions is equivalent to the case of point-like junctions with slightly longer waveguides. The latter is in line with our theoretical model. We also make sure that the effective length of each waveguide is identical.
 
\begin{figure}
    \centering
    \includegraphics[width=\columnwidth]{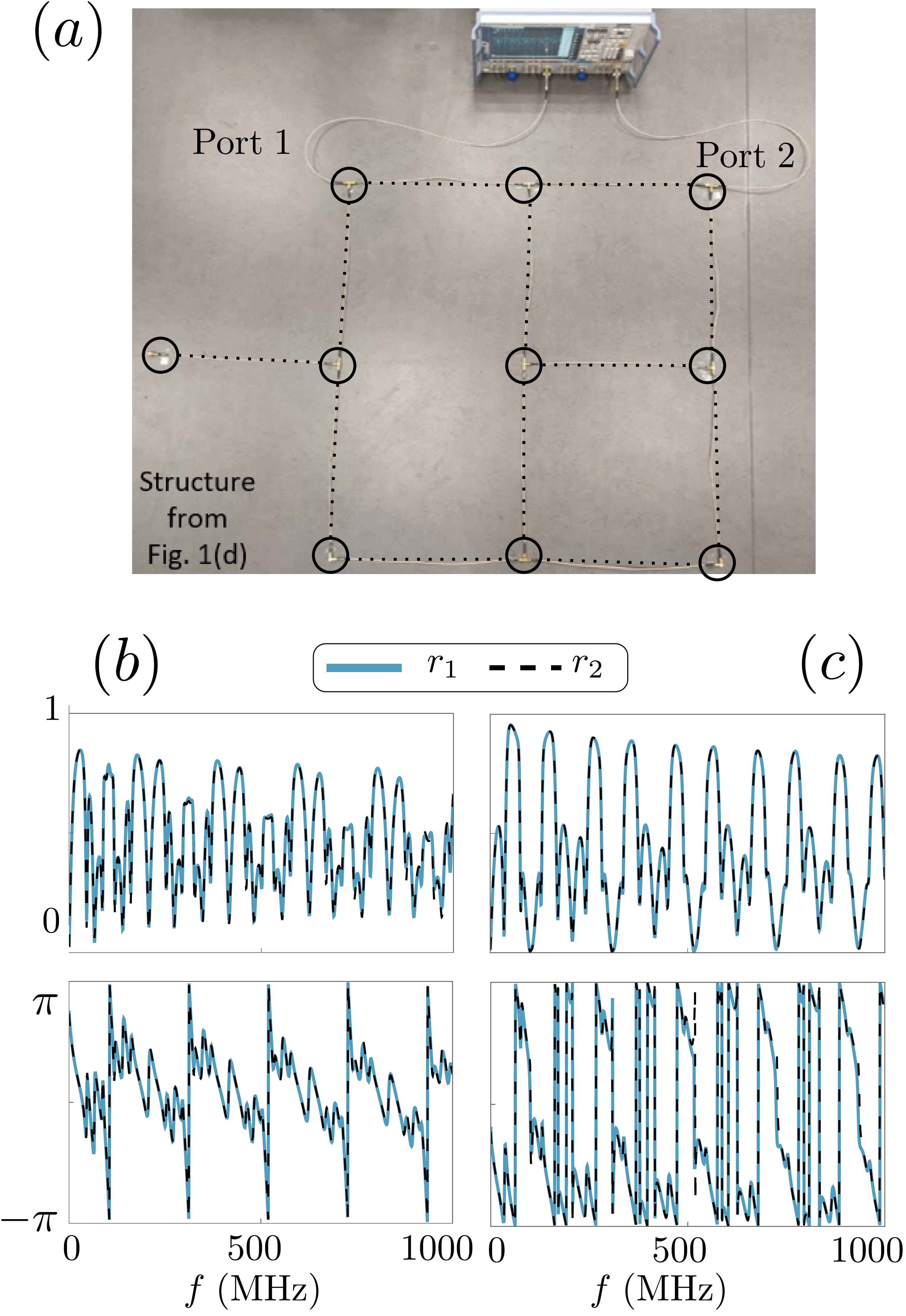}
    \caption{
    Experimental scattering measurements on microwave transmission-line networks with latent symmetry. (a) Photography of the experimental setup for the network from \cref{fig:firstNetwork}(d). (b) and (c) Magnitude and phase of the measured reflection coefficients for the networks from \cref{fig:firstNetwork}(d) and \cref{fig:moreNetworks}(a), respectively.}
    \label{fig:MicrowaveExperiments}
\end{figure}

 We observe in \cref{fig:MicrowaveExperiments} excellent agreement of the complex-valued reflection coefficients (i.e., in terms of both magnitude and phase) with the expected equireflection condition at all frequencies. Moreover, a clear periodicity of the reflection spectrum is visible. Furthermore, it is apparent that the absorption strength increases monotonously with frequency, as expected.

 \section{Powers of the scattering matrix and perfect absorption} \label{sec:powersOfTheSMatrix}
	
	Although two-port systems provide an intuitive and simple setup to study wave scattering, many applications are actually based on $N$-port configurations (circulators, splitters, multiplexers etc.).  In view of these implementations, in this Section, we study latent symmetry and generalisations of it \textit{directly on the scattering matrix itself} and not on the underlying structure. In such a way the results obtained here do not depend on the specific physical system and are rather general for any N-port scattering matrix $S$.

We first note that any reciprocal scattering matrix is symmetric $S = S^{T}$. Let us then consider a case where this matrix is also latently symmetry; that is, it fulfills
		\begin{equation} \label{eq:generalCospectralDefinition}
	\left(S^{\matPowerIndex} \right)_{\su,\su} = \left(S^{\matPowerIndex} \right)_{\sv,\sv}, \quad \forall \quad m=1,\ldots{},N
		\end{equation}
  for some $\su$ and $\sv$. Note that this relation for $\matPowerIndex=1$ implies that ports $\su$ and $\sv$ have the same reflection coefficient.
 We emphasize that the connection between matrix powers, cospectrality and eigenvectors (presented in the last section) is valid for the scattering matrix as well. 
 Thus, if \cref{eq:generalCospectralDefinition} is valid, the eigenvectors of the S matrix will have parity between the elements $\su$ and $\sv$.
However, the eigenvectors of the scattering matrix, except for particular cases (as we show below) are not so useful. 
We remind the reader that \cref{eq:generalCospectralDefinition} also implies cospectrality between $S \setminus \su$ and $S\setminus \sv$ but gives no information about the eigenvalues of $S$ itself. To bypass this, let us go one step further and generalise \cref{eq:generalCospectralDefinition} by introducing the following relation
    \begin{equation} \label{eq:scaledCospectrality}
        \left(S^{\matPowerIndex} \right)_{\su,\su} =c \cdot \left(S^{\matPowerIndex} \right)_{\sv,\sv} \quad \forall \quad \matPowerIndex=1,\ldots{},N \, .
    \end{equation}
We call this relation, \cref{eq:scaledCospectrality}, \emph{scaled cospectrality} of a matrix $S$ with a scaling factor $c$. Interestingly, this rather simple generalisation has direct impact on the eigenvalues of the matrix $S$:
whenever $c\ne 1$, scaled cospectrality enforces the existence of one zero eigenvalue (more zeros might appear accidentally). 
To show why this is the case, let us start by realizing that, from the Cayley-Hamilton theorem, we know that $S^N$ can be written as a polynomial in the first $N-1$ powers of $S$, that is, 
		\begin{equation} \label{eq:charPolynomial}
        \sMat^{N} = - \sum_{\matPowerIndex=0}^{N-1} a_{\matPowerIndex} \sMat^{\matPowerIndex} \, ,
		\end{equation}
  where $S^{0} \equiv I$ is the identity matrix and where the coefficients $a_{m}$ are taken from the characteristic polynomial of $S$, given by
        $P(x) = \mathrm{det}\left( x \, I - S\right) = \sum_{\matPowerIndex=0}^{N} a_{\matPowerIndex} x^{\matPowerIndex}$.
    Now, since the matrix $S$ fulfills $\component{\sMat^{N}}{\su}{\su} = c \component{\sMat^{N}}{\sv}{\sv}$, it follows from \cref{eq:charPolynomial} that
    \begin{equation}
        -a_{0} I_{\su,\su} - \sum_{\matPowerIndex=1}^{N-1} a_{\matPowerIndex} \component{\sMat^{\matPowerIndex}}{\su}{\su} = c \left(-a_{0} I_{\sv,\sv} - \sum_{\matPowerIndex=1}^{N-1} a_{\matPowerIndex} \component{\sMat^{\matPowerIndex}}{\sv}{\sv} \right) 
    \end{equation}
    from which we get
    \begin{equation}
        a_{0} \component{I}{\su}{\su} = c \, a_{0} \component{I}{\sv}{\sv} \, .
    \end{equation}
    It follows that $a_{0} = 0$ whenever $c\ne 1$ and since $a_{0}$ is proportional to the determinant of $\sMat$, we see that $\sMat$ has to have a vanishing eigenvalue.
The existence of a zero eigenvalue of $S$ directly implies that there must be a monochromatic adapted wavefront that can be injected at the real frequency at which $S$ has the zero eigenvalue such that it is perfectly absorbed within the system~\cite{chen2020perfect}. This CPA requires the system to have a finite amount of absorption loss and hence a sub-unitary $S$ matrix. Indeed, in the absence of any absorption loss, the $S$ matrix would be unitary and the magnitude of its determinant would be unity, implying that $a_0$ cannot be zero because $a_0$ is proportional to det($S$). 

Let us remark for the interested reader that in the literature there is also another generalization of cospectrality, the so-called \emph{fractional cospectrality} \cite{Chan2020AQFundamentalsFractionalRevivalGraphs}, which, however, does not enforce the presence of a zero eigenvalue.

In this section, we have so far not explicitly stated the frequency dependence of $S$. However, a scattering system can in general only fulfill the conditions for scaled cospectrality at discrete frequencies, which is hence in line with the fact that the zeros of $S$ only occur at discrete (potentially complex) frequencies. In other words, if one optimizes the scattering system such that it has scaled cospectrality (and thus CPA) at some frequency $f_0$, there is no reason to expect scaled cospectrality at any frequency other than $f_0$.

In the remainder of this section, we will show that systems with scaled cospectrality allow us to achieve CPA with wavefronts that have a customized imbalance in terms of the weights of different channels. Before explaining the origin of this feature in detail, let us briefly contextualize this feature within the recent literature on achieving CPA.
The $S$ matrix of a generic (arbitrarily complex) scattering system does not necessarily have a zero eigenvalue, but can be tuned to have one~\cite{Fyodorov2017JPAMT5030LT01DistributionZerosSmatrixChaotic,KottosPRLCPA,pichler2019random,chen2020perfect}. With sufficient tunable degrees of freedom, the real frequency at which the zero eigenvalue occurs can be controlled, too~\cite{f2020perfect,frazierCPA,delHougne2021LPR152000471OnDemandCoherentPerfectAbsorption}. However, in structures without symmetry the required CPA wavefront is in general highly asymmetrical and difficult to generate~\footnote{See Supplementary Note 4 in Ref.~\cite{Sol2023SA9eadf0323ReflectionlessProgrammableSignalRouters} for a detailed analysis in terms of entropy and participation number of CPA wavefronts in a chaotic cavity}. This makes it more challenging to realize the envisaged use of CPA to interferometrically control light with light, and hence without any non-linearity~\cite{Zhang2012LSA1e18ControllingLightwithlightNonlinearity,baranov2017coherent}. 
A generic way to impose CPA at a desired frequency and with an arbitary CPA wavefront was demonstrated in Ref.~\cite{delPRL2021} by tuning a massively parametrized chaotic cavity. Here, we provide an alternative route through scaled cospectrality to obtain structures featuring CPA whose wavefront has a prescribed imbalance. Thereby, only a weak control wave can modulate a strong signal, provided that the necessary phase and amplitude relation between the two is respected.

\begin{figure}[h!] 
	\centering
	\includegraphics[max width=\linewidth]{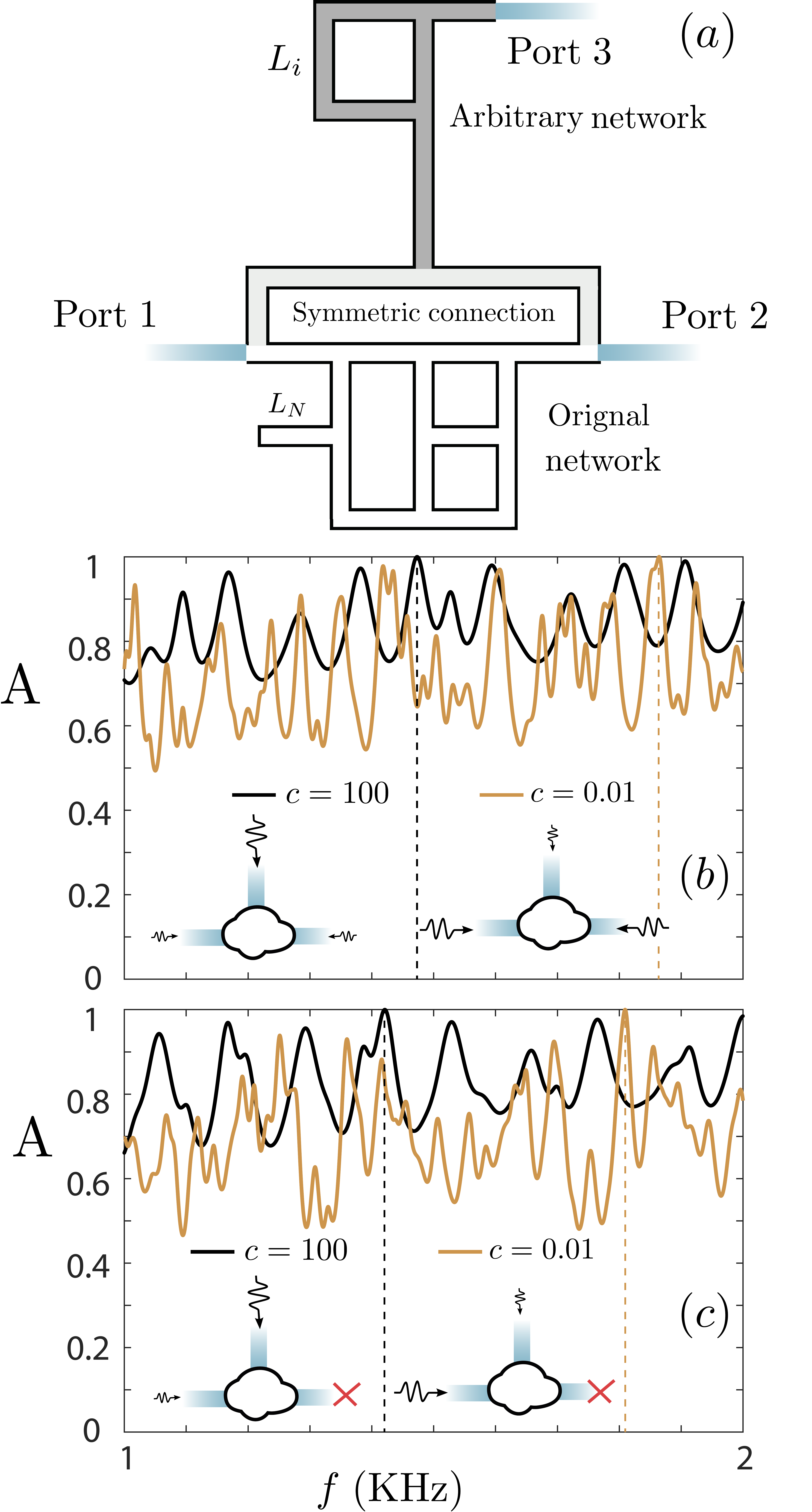}
 \caption{(a) A symmetric extension of the $2$-port setup shown in \cref{fig:firstNetwork} (d). For (b) and (c), we finetune the lengths of the acoustic waveguides of the network (a) to fulfill scaled cospectrality between ports $1$ and $3$ with the scattering matrix fulfilling either \cref{eq:solution2} [for (b)] or \cref{eq:solution1} [for (c)]. For each (b) and (c), we show the frequency-dependent absorption $A$ of the network for the two different choices $c=100$ and $c=1/100$.}
	\label{fig:extraPort}
	\end{figure}

We now illustrate the above results with an  implementation of the scaled cospectrality using a particular example of a 3-port acoustic network with lossy acoustic waveguides (see \cref{app:lossySystems} for details).
For simplicity we choose to work with an extension of the 2-port system of the previous section by symmetrically coupling an arbitrary structure to $a$ and $b$ as shown in \cref{fig:extraPort} (a). The symmetric connection and the added structure are highlighted by light gray and dark gray respectively. 

The two latently symmetric points are symmetrically connected to ports 1 and 2, and thus the scattering matrix of the 3-port network reads
	\begin{equation} \label{eq:extendedScatteringMatrix}
    S = \begin{pmatrix}
    r & t & t' \\
    t & r & t'\\
    t' & t' & r'
     \end{pmatrix}.
    \end{equation}
    with $r,r'$ denoting reflection coefficients, and with $t,t'$ transmission coefficients.
  Evidently the proposed design ensures the equireflectionality ($r$) of the two symmetric ports 1 and 2 while an additional reflection ($r'$) from port 3 is introduced. Due to reciprocity and  symmetry we are thus left with only two different transmission coefficients $t$ and $t'$.
    For this particular matrix we impose scaled cospectrality between ports $1$ (or equivalently $2$) and $3$ by demanding that it satisfies \cref{eq:scaledCospectrality}.
Note that for $\matPowerIndex=1$ we simply get a scaling between the reflections, $r = c r'$. Then, solving for the higher powers we obtain two possible solutions which are discussed in the following.
\paragraph*{CPA without any zero-input channel:}

One solution requires 
	\begin{equation} \label{eq:solution2}
	 t = -r'/2, \mbox{ and } t' = r' \sqrt{2c-1}/2 \, .
	\end{equation}
To illustrate the effects of the zero eigenvalue of $S$ and its relation to CPA, we define the following eigenvalue problem
\begin{equation} \label{eq:Sceig}
    	S\mathbf{x}=\lambda \mathbf{x}.
    \end{equation} 
When  $S$ satisfies \cref{eq:solution2}, the corresponding eigenvalues of the scattering matrix are
$\lambda_1=(0,(c+1/2)r',(c+1/2)r')$. As expected by our definition of the generalised cospectrality, the scattering matrix acquires a zero eigenvalue. This zero eigenvalue implies that if one uses as an input the corresponding eigenvector, the output is \textit{zero} and the input wavefront is perfectly absorbed. In fact, here the corresponding eigenvector is 
\begin{equation}
    \mathbf{x}_1=(-1/\sqrt{2c-1},-1/\sqrt{2c-1},1)^T \, .
\end{equation}
Importantly, the imbalance of the input wavefront is solely controlled by the scaling factor $c$ which we can prescribe.

In order to validate our results, we once again consider the case of airborne acoustic waveguides including viscothermal losses. The elements of the $S$ matrix in such a case can be calculated using different techniques such as the star product (see \cref{app:optimProcedure} for details). 
We optimize the lengths of several parts of the network (while maintaining the symmetry between ports 1 and 2) such that \cref{eq:solution2} is satisfied for a predefined value of $c$ and a suitable frequency $\optimizedFreq{}_1$. To illustrate our result, we assume an input to the 3-port network in the form of a vector $\mathbf{p}_{\rm{in}} = \mathbf{x}_{1}$ and thus the output vector is naturally given by $\mathbf{p}_ {\rm{out}} = S \, \mathbf{p}_{\rm{in}}$. Using the output we can calculate 
absorption from such networks as 
\begin{equation}
    A \equiv 1 - \frac{\left| \mathbf{p}_{\rm{out}}\right|^{2} }{\left| \mathbf{p}_{\rm{in}}\right|^{2}}.
\end{equation}
The absorption for two different networks is shown in \cref{fig:extraPort}(b). The black thick (orange thin) line corresponds to a network satisfying  \cref{eq:solution2} with $c=100$ ($c=0.01$). The details of the geometry of the obtained networks are given in \cref{app:optimProcedure}.
According to our design, at the prescribed frequency $\optimizedFreq{}_1$, CPA of the input wave is achieved, indicated by $A=1$.
We additionally observe that with relatively small variations of the lengths of the waveguides (see \cref{app:optimProcedure}), we achieve CPA for a very different input vector of an amplitude $10$ times larger (smaller) for port $1/2$ compared to port $3$.

\paragraph*{CPA with one zero-input channel:}
The other possible solution for $S$ to acquire scaled cospectrality is when 
 \begin{equation} \label{eq:solution1}
    	t = r= cr' \mbox{ and } t' = \sqrt{c}r' \, ,
    \end{equation} 
with the eigenvalues of $S$ then being $\lambda_2=(0,0,(2c+1)r')$. 
The corresponding eigenvectors of the two zero eigenvalues are $\mathbf{x}^{(1)}_2=(1,-1,0)^{T}$ and $\mathbf{x}^{(2)}_2=(1,1,-2\sqrt{c} )^{T}$.
This case is even more interesting since it features a two-fold degenerate zero eigenvalue~\cite{FanDegenCPA}. Consequently, for any choice of $\alpha,\beta$, inputs of the form $\alpha \mathbf{x}^{(1)}_2 + \beta \mathbf{x}^{(2)}_2$ will be perfectly absorbed. Among all possible inputs, one can find the following highly asymmetric one:
\begin{equation}
    \mathbf{x}_2=(-1,0,\sqrt{c})^T \, .
\end{equation}
Thus, the 3-port network is able to completely absorb waves with non-zero inputs on only two of its ports and with a relative input between the two active ports prescribed by the scale factor $c$. We have constructed two such networks using $c=100$ and $c=0.01$, satisfying \cref{eq:solution1} at a prescribed frequency $\optimizedFreq{}_2$ and the corresponding absorption is shown in \cref{fig:extraPort}(c) using the input $\mathbf{p}_{in} = \mathbf{x}_{2}$. The networks feature CPA ($A=1$) at the desired frequency $\optimizedFreq{}_2$ with non-zero inputs in only two of the three channels. On top of that, the imbalance of the CPA wavefront introduced by the scale factor $c$ results in an almost single sided (or one port) CPA.
	
	\section{Conclusions} \label{sec:Conclusions}
    We have studied the scattering properties of various asymmetric waveguide networks and demonstrated that a certain family of networks possesses the same scattering properties as mirror-symmetric ones, i.e. they show equireflectionality. This counterintuitive property of the asymmetric networks stems from a hidden mirror symmetry called latent symmetry. We have validated this finding numerically for acoustic waveguide networks and experimentally for microwave transmission-line networks.

While latent symmetry is mathematically equivalent to certain relations obeyed by the matrix powers, here we generalized these relations and applied them to a generic scattering matrix of an $N$-port system. The new relations, named scaled cospectrality, were then used to construct networks featuring CPA. Specifically, a scaling factor was used to design systems able to completely absorb wavefronts with prescribed imbalance, a capability that can enable the control of light with a very weak coherent control signal. Overall, our work demonstrates that scattering problems may greatly profit by the notion of latent symmetry and of matrix power relations in general. 
 
	\begin{acknowledgments}
		The authors are thankful to M. Pyzh and V. Pagneux for valuable discussions.
	\end{acknowledgments}
	\clearpage

	\onecolumngrid
	\appendix
 
	\section{Eigenvectors of symmetric scaled cospectral matrices} \label{app:EigenvectorProof}
	
        Let $\sMat = \sMat^{T} \in \mathbb{C}^{N \times N}$ be a complex-symmetric matrix, and let us assume that $\component{\sMat^{\matPowerIndex}}{\su}{\su} = c\cdot \component{\sMat^{\matPowerIndex}}{\sv}{\sv}$ for all $\matPowerIndex>0$. It follows that 
        \begin{equation}
			\component{e^{i \sMat t} - 1}{\su}{\su} = c \cdot \component{e^{i \sMat t} - 1}{\sv}{\sv}.
		\end{equation}

        We assume that $\sMat$ has no degenerate eigenvalues, which in particular implies that $S$ is diagonalizable. Then, since $\sMat = \sMat^{T}$, one can normalize its eigenvectors $\ket{\phi_{i}}$ such that $\braket{\phi_{j}^{*}|\phi_{i}} = \delta_{i,j}$ with the star denoting the complex conjugate. As a consequence, we have $1 = \sum_{i} \ket{\phi_{i}}\bra{\phi_{i}^{*}}$.
		Equipped with this identity, we get
		\begin{align}
			\component{e^{i \sMat t} - 1}{\su}{\su} =& \sum_{i,j} \braket{\su|\phi_{i}} \braket{\phi_{i}^{*}|e^{i \sMat t}|\phi_{j}} \braket{\phi_{j}^{*}|\su} - 1 \\
			=& \sum_{i} \braket{\su|\phi_{i}} e^{i \lambda_{i} t} \braket{\phi_{i}^{*}|\su} - 1 \\
			=& c \cdot \left(\sum_{i} \braket{\sv|\phi_{i}} e^{i \lambda_{i} t} \braket{\phi_{i}^{*}|\sv} - 1 \right)
		\end{align}
		with $\lambda_{i}$ being the eigenvalue of $\ket{\phi_{i}}$. Since the complex exponentials are linearly independent, one can evaluate the above equation independently for each distinct $\lambda_{i}$, so that it automatically follows that 
		\begin{align}
			\braket{\su|\phi_{i}} \braket{\phi_{i}^{*}|\su} &= c \cdot \braket{\sv|\phi_{i}} \braket{\phi_{i}^{*}|\sv},\quad \lambda_{i} \ne 0 \\
			\braket{\su|\phi_{0}} \braket{\phi_{0}^{*}|\su} &= c \cdot \braket{\sv|\phi_{0}} \braket{\phi_{0}^{*}|\sv} -c + 1,\quad \lambda_{0} = 0.
		\end{align}
		Since $\braket{\phi_{j}^{*}|\su} = \braket{\su|\phi_{j}}$ for \emph{all} eigenvectors, we obtain
		\begin{align}
			\braket{\su|\phi_{i}} &= \pm \sqrt{c} \cdot \braket{\sv|\phi_{i}}, \quad \lambda_{i} \ne 0 \\
			\left(\braket{\su|\phi_{0}} \right)^2 &= c \cdot \left(\braket{\sv|\phi_{0}} \right)^2 - c +1, \quad \lambda_{0} = 0.
		\end{align}

	\section{Equireflectionality in a lossy system} \label{app:lossySystems}
	In the main text of this work, we considered the case of latent reflection symmetry in a system without losses, and we mentioned that the point-wise parity of eigenmodes remains valid for the case of losses in the one-dimensional waveguides described by a complex velocity $c = c_{r} + i c_{i}$. In the following, we give the justification for this statement.

	To this end, let us assume that we have a network which, in the absence of losses, has a latent reflection symmetry between two junctions $\su,\sv$. As stated in the main text, for low-enough frequencies, the eigenmodes of this network can be found from the generalized eigenvalue problem
	\begin{equation} \label{app:gEVP}
	    A\, \boldsymbol{\phi} = \cos(k L) B\, \boldsymbol{\phi}
	\end{equation}
	where $\boldsymbol{\phi}$ contains the pressure of the eigenmode $\phi$ at the $N$ junctions of the network, and with $k = \omega/c = \omega/c_{r} \equiv k_{r}$. The eigenvalues $\cos(k L)$ of this problem are completely real and due to latent symmetry, all eigenmodes fulfill $p_{\su} = \pm p_{\sv}$.

    Let us then take the usual route for introducing losses: That is, we start from the lossless case---as described through \cref{app:gEVP}---and then let the velocity $c =c_{r} + i c_{i}$ (due to, in particular, thermo-viscous boundary at the surface of the one-dimensional waveguides \cite{kosten1949sound}). Obviously, this changes the relation between $k$ and $\omega$, but it does not change the eigenvalues $\cos(k L)$ or the eigenvectors $\boldsymbol{\phi}$. In other words, each eigenvector $\boldsymbol{\phi}$ with eigenvalue $\cos(k L)$ of the lossless system will still be an eigenvector of the lossy system, with unchanged eigenvalue $\cos(k L)$. In particular, point-wise parity of eigenmodes is preserved.
    What changes is the frequency $\omega$ corresponding to $\cos(k L)$: Since $c$ is complex while $k$ is real (as imposed from the fact that $\cos(k L)$ is real), $\omega$ is in general complex as well, with the imaginary part being related to the lifetime of this lossy eigenmode. 
 
	\section{Using the matrix power relations for deriving a better understanding of latently symmetric waveguide networks} \label{app:matrixPowerRelationsNeighbours}
    The aim of this section is to showcase a set of intuitive and easily interpretable equivalent conditions that the relations \cref{eq:matrixPowerRelations}---latent symmetry, that is---impose on a waveguide network. These conditions can be derived by analyzing \cref{eq:matrixPowerRelations} order by order and subsequently using the relation $\ham = B^{-1/2} A B^{-1/2}$ to derive conditions on the matrices $A,B$ which are directly describing the underlying waveguide network. We restrict ourselves to the first few orders of \cref{eq:matrixPowerRelations}, for which the corresponding equivalent conditions have been derived in \cite{Rontgen2023PRL130077201HiddenSymmetriesAcousticWave}; we repeat them here for self-containedness of the present manuscript.

    The conditions are then as follows.
    Firstly, $\su$ and $\sv$ have to have the same number of neighbors. Moreover, in the special case where the number of next-neighbors of $\su,\sv$ is equal to unity, (i) \cref{eq:matrixPowerRelations} holds for $\matPowerIndex=2$ if and only if $\su,\sv$ have the same number of next-neighbors, and (ii) \cref{eq:matrixPowerRelations} holds for $\matPowerIndex=3$ if and only if
    \begin{equation} \label{eq:degreeRelation}
        \sum_{i \in \mathcal{N}^2(\su)} \frac{1}{|\mathcal{N}(i)|} = \sum_{i \in \mathcal{N}^2(\sv)} \frac{1}{|\mathcal{N}(i)|} \, .
    \end{equation}
    In this relation, $\mathcal{N}^{2}(i)$ denotes the set of next-neighbors of $i$, and $|\mathcal{N}(i)|$ denotes the degree of site $i$, that is, the number of neighbors of $i$.

    Equipped with the above, we can now analyse the difference between the setups shown in \cref{fig:firstNetwork} (c) and \cref{fig:firstNetwork} (d) in more detail. To this end, let us close these two systems on the entry-points of the two ports, and call these points $\su$ (on the left-hand side of the setup) and $\sv$ (on the right-hand ride of the setup); compare  \cref{fig:Mapping_To_A_Graph}. 

    For the asymmetric network of \cref{fig:firstNetwork} (c), the two next-neighbors of $\su$ have degrees $3$ and $2$, while the two next-neighbors of $\sv$ both have degree $3$.
    Thus, the equation \cref{eq:degreeRelation} is not fulfilled  and, as a consequence, the relations \cref{eq:matrixPowerRelations} are not fulfilled for $\matPowerIndex=3$. 
    On the other hand, the setup of \cref{fig:firstNetwork} (d) features an additional waveguide on the left, which equalized the two sides of \cref{eq:degreeRelation}.
    
    \section{Optimization procedure and lengths of the optimized structures} \label{app:optimProcedure}
\begin{figure}[htb] 
	\centering
	\includegraphics[max width=0.5\linewidth]{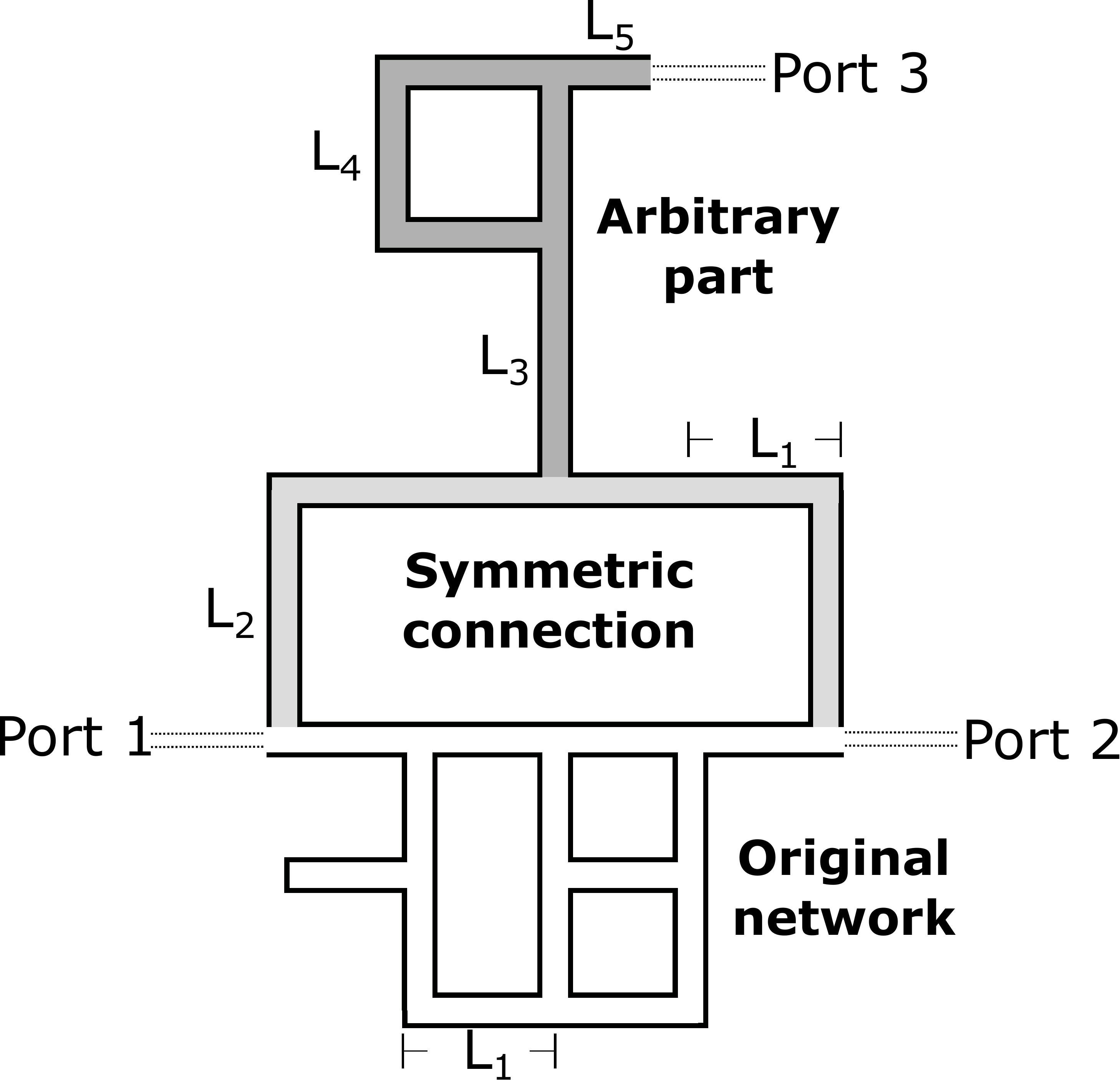}
	\caption{Reference figure showing the assignment of the waveguide lengths $L_1$, \ldots , $L_5$.
	}
	\label{fig:optimizationSketch}
	\end{figure}

    In order to design a system whose scattering matrix has the form of \cref{eq:extendedScatteringMatrix} and which fulfills either \cref{eq:solution1} or \cref{eq:solution2}, we proceed as follows: We started with the system depicted in \cref{fig:extraPort} (a), whose scattering matrix has the structure of \cref{eq:extendedScatteringMatrix}. We then optimized the lengths $L_{1},\ldots{},L_{5}$ (see \cref{fig:optimizationSketch}) and the frequency $f$ in the range between $1000$ and $2000$ Hz such that \cref{eq:solution1,eq:solution2} are fulfilled, respectively, at a frequency $\optimizedFreq{}$. Technically, we modeled the system as a network of ideal waveguides of length $L_{m}$, each with a transmission coefficient
    \begin{equation} \label{eq:transmissionInEachWaveguide}
        t_{m} = exp(-i k L_{m})
    \end{equation}
    with \begin{equation} \label{eq:lossesEquation}
        k = \frac{2\pi f}{c_s} + (1+i) \, \alpha \, \frac{\sqrt{f}}{R};
    \end{equation}
    where $\alpha$ is the loss coefficient (see below) \cite{kosten1949sound},  $R = 1\, cm$ the waveguide diameter, with $c_{s}=343\,m/s$ the velocity of sound and $f$ being the frequency. Using the continuity of the pressure and the conservation of the flux at each junction, we then obtain the scattering matrix $S$, depending only on the frequency $f$ and the lengths $L_{m}$.

    The optimized lengths (in meters, rounded to 4 digits after the decimal point) are shown in the table below. The cases a and b correspond to the scattering matrix fulfilling \cref{eq:solution2} and \cref{eq:solution1} of the main text, respectively.

    \begin{center}
        \begin{tabular}{|c|c|c|c|c|}
	\hline
	& case a, $c=1/100$ & case a, $c=100$ & case b, $c=1/100$ & case b, $c=100$ \\
	\hline
	$L_1$ & $0.7976$ & $0.1021$ & $0.7792$ & $0.1326$ \\
	\hline
	   $L_2$ & $0.6398$ & $0.6003$ & $0.3944$ & $0.7917$ \\
	\hline
	$L_3$ & $0.1070$ & $0.3201$ & $0.4018$ & $0.6452$ \\
	\hline
	$L_4$ & $0.7497$ & $0.8000$ & $0.3357$ & $0.7232$ \\
	\hline
	$L_5$ & $0.7837$  & $0.7045$ & $0.3289$ & $0.5720$ \\
	\hline
\end{tabular}
    \end{center}

    \clearpage
        \section{Confirming the reciprocity of the transmission line networks}
\begin{figure}[htb!]
    \centering
    \includegraphics[max width=0.5\columnwidth]{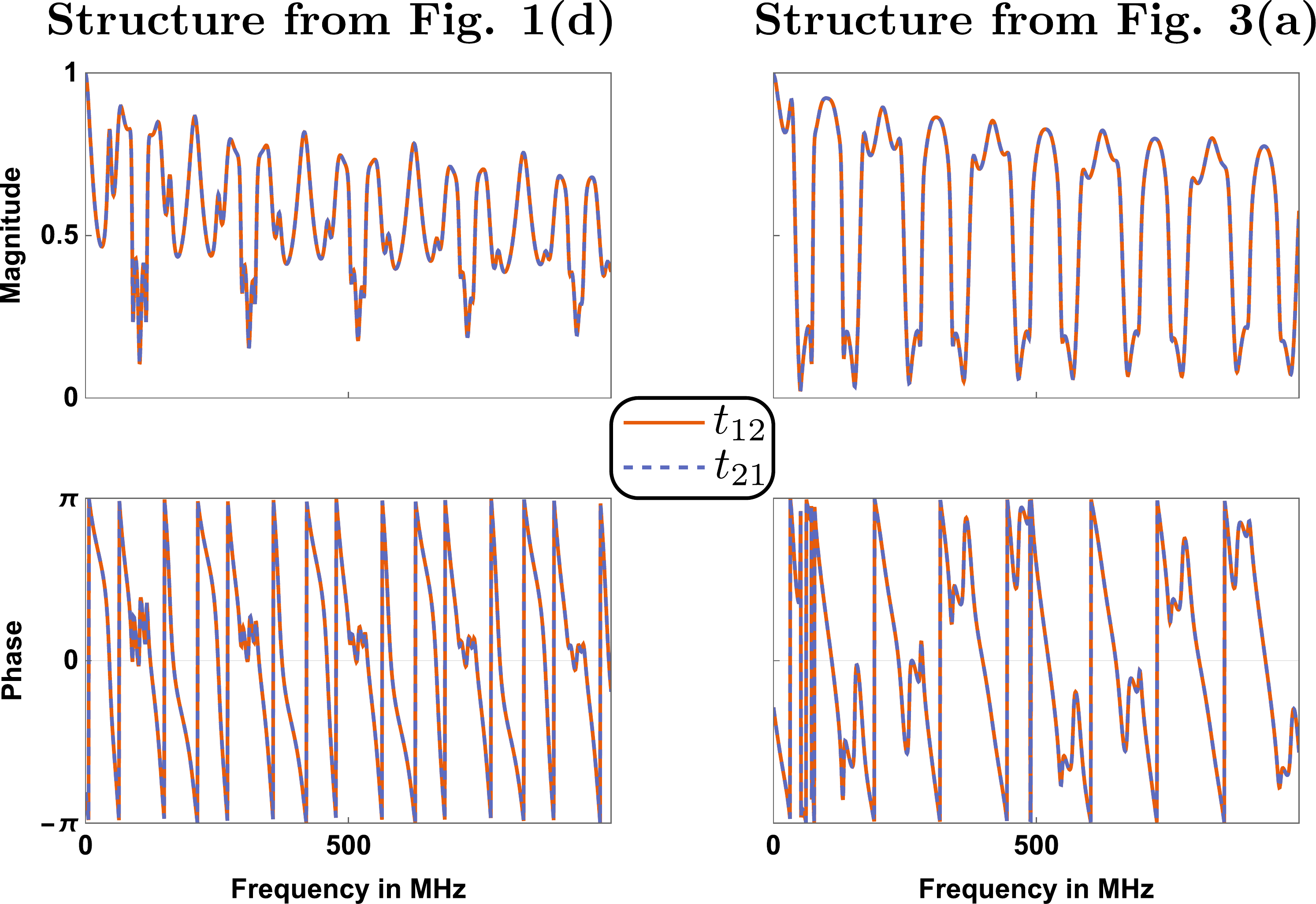}
    \caption{Experimentally measured transmission coefficients $t_{ij}$ from port $i$ to port $j$ for the microwave transmission-line networks from \cref{fig:firstNetwork}(d) (left) and \cref{fig:moreNetworks}(a) (right).}
    \label{fig:MicrowaveExperimentsTransmission}
\end{figure}

        In the main text, we have already shown the reflection coefficients for the transmission line measurements of the networks shown in \cref{fig:firstNetwork}(d) and \cref{fig:moreNetworks}(a).
        In \cref{fig:MicrowaveExperimentsTransmission}, we also show the transmission coefficients, that is, the matrix elements $S_{1,2}$ and $S_{2,1}$ of the scattering matrix $S$. As can be seen, the networks are reciprocal in their scattering properties, that is, they fulfill $S_{1,2}=S_{2,1}$.
    
\clearpage

\end{document}